\documentclass[aps,twocolumn,grouedaddress,superscriptaddress,a4paper,showkeys]{revtex4-1}
\usepackage{amssymb}
\usepackage{amsmath}
\usepackage[english]{babel}
\usepackage{graphicx}

\begin{document}

\title{Quantum efficiencies in finite disordered networks connected by  
many-body interactions}

\author{Adrian Ortega}
\email{adrianortega@fis.unam.mx}
\affiliation{Instituto de Ciencias F{\'i}sicas, 
Universidad Nacional Aut{\'o}noma de M\'{e}xico, 
62210 Cuernavaca, M\'{e}xico}

\author{Manan Vyas}
\email{manan@fis.unam.mx}
\affiliation{Instituto de Ciencias F{\'i}sicas, 
Universidad Nacional Aut{\'o}noma de M\'{e}xico, 
62210 Cuernavaca, M\'{e}xico}

\author{Luis Benet}
\thanks{Corresponding author} 
\email{benet@fis.unam.mx}
\affiliation{Instituto de Ciencias F{\'i}sicas, 
Universidad Nacional Aut{\'o}noma de M\'{e}xico, 
62210 Cuernavaca, M\'{e}xico}
\affiliation{Centro Internacional de Ciencias, Cuernavaca, M\'{e}xico}

\begin{abstract}

  The quantum efficiency in the transfer of an initial excitation in disordered 
  finite networks, modeled by the $k$-body embedded Gaussian ensembles of random 
  matrices, is studied for bosons and fermions. The influence of the presence or absence of 
  time-reversal symmetry and centrosymmetry/centrohermiticity are addressed. For
  bosons and fermions, the best efficiencies of the realizations of the ensemble 
  are dramatically enhanced when centrosymmetry (centrohermiticity) is imposed.
  For few bosons distributed in two single-particle levels this permits perfect state 
  transfer for almost all realizations when one-particle interactions are considered. For 
  fermionic systems the enhancement is found to be maximal for cases when all but
  one single particle levels are occupied.

\end{abstract}


\maketitle

\section{Introduction}

An important question on complex quantum systems, which remains open to a large extent,
addresses the conditions to have robust efficient transport of excitations across a disordered 
finite network~\cite{lee_science,tor_pre,scholes_review}. Whereas it is well known how to 
define a Hamiltonian system where perfect transport is 
obtained~\cite{christandl_prl, christandl_pra}, 
the fact that such system requires the precise specification of a large number of parameters 
makes it difficult to achieve in practice. This is the sense of robustness above: 
statistical changes 
in the parameters should lead to small fluctuations that preserve good efficiencies, instead of 
large transmission fluctuations. Clearly, the number of control parameters should be as small 
as possible.

Inspired by Ref.~\cite{mattia_prl}, we study here the distribution of the transport 
efficiencies of an initial localized excitation in a disordered network of $l$ sites,
which is modeled by random Hamiltonian that includes many-body interactions, considering 
both fermions and bosons; to the best of our knowledge, quantum efficiencies
of this type of disordered
networks have not been considered. In general, this question is of interest in a 
variety of fields, including understanding photosynthetic light-harvesting 
complexes~\cite{1997PhysToday}, such as the Fenna-Matthews-Olson (FMO)
complex~\cite{fennamatthews_nature}, or in quantum communication protocols 
across quantum spin chains~\cite{bose_prl, 2005NJP-BB, bose_review, kay_review}. 
In either case, a model with few-body interactions seems realistic.

The transport efficiency from an input state $|{\rm in}\rangle$ to an output state 
$|{\rm out}\rangle$, is quantified as the maximum transition probability achieved 
among theses states within a time interval $[0,T]$. The transport 
efficiency is defined as~\cite{mattia_prl}
\begin{equation}
\mathcal{P}_{{\rm in},{\rm out}} = \max_{[0,T]} 
  |\langle {\rm out}| U(t)|{\rm in}\rangle|^2.
\label{eq1}
\end{equation}
Here, $U(t)$ is the unitary quantum evolution associated with the Hamiltonian of the 
system and $T$ is a reasonable time scale ($\hbar=1$). 
The system is said to have perfect state transfer (PST) when 
$\mathcal{P}_{{\rm in},{\rm out}} = 1$~\cite{christandl_prl}. 
The nodes of the network are the basis states of the 
Hilbert space where the initial excitation is localized. We define the Hilbert
space by distributing $n$ spinless particles (bosons or fermions) in $l$ single-particle 
states. The $n$-body Hamiltonian that we consider consists of a random $k$-body interaction 
among the $n$-particles states, with $1\leq k\leq n$; the non-zero Hamiltonian matrix 
elements are related to the links of the network. This matrix model is known as 
the (bosonic or fermionic) $k$-body embedded Gaussian ensemble of random 
matrices~\cite{2003JPA-BW,kotabook}. 

The motivation for choosing this matrix model comes from the fact that this ensemble
displays correlations among the matrix elements~\cite{2001AnnPhys-BRW,2002AnnPhys}, in
the sense that the number of independent random variables is usually smaller than 
the number of independent links of the network. In addition, the bosonic ensemble for 
the specific case where the bosons are distributed in two single-particle levels 
displays a systematic appearance of doublets in the spectrum~\cite{saul}. Thus, one of 
the design principles required in Ref.~\cite{mattia_prl}, the
existence of a dominant doublet, is automatic fulfilled. The extension to fermions 
arises naturally, and is also motivated by the transport processes within
the FMO system~\cite{1997PhysToday}. 

As in previous studies~\cite{2014NJP-ZMWB}, centrosymmetry~\cite{cantoni} 
is considered and 
we find that it increases dramatically the best efficiencies among certain pairs of 
distinct sites, in comparison to a non-centrosymmetric interaction; we also study in
this paper the case where time-reversal symmetry is broken, and the corresponding
case including centrohermiticity~\cite{1990SIAMMatAnalApp}. Here, as usual, 
the benchmark for good efficiencies is $95\%$. In particular, for $k=1$ and few 
bosons distributed in two single-particle states, we find that PST is obtained for a 
large fraction of the best efficiencies of the ensemble
when centrosymmetry or centrohermiticity are included; 
for larger values of $k$ the best efficiencies are reduced. 
In the case of fermions and including centrosymmetry or centrohermiticity, we show 
that below half-filling ($n/l < 1/2$, where $n$ is the number of fermions and $l$ the
number of single-particle states) the best 
efficiencies are obtained for $k=1$; above half-filling, our numerics
indicate that the best efficiencies 
correspond to $k\sim l/2$ for $n=l-1$. In both cases, the benchmark is 
achieved with non-zero probability.

The paper is organized as follows: in Sect.~\ref{sec2} we define the matrix model
considered and centrosymmetry, which is introduced at the one-particle level and then 
extended to the $k$- and $n$-particle spaces. In Sect.~\ref{sec3} we describe the 
results for the distribution of the best efficiencies of each realization of the 
ensemble for bosons occupying two-single particle states, with or without the 
addition of centrosymmetry or centrohermiticity; Sect.~\ref{sec4} analyzes the 
fermionic case along the same lines. Finally, in Sect.~\ref{sec5} we present 
the conclusions of this work.

\section{The embedded Gaussian ensembles and centrosymmetry / centrohermiticity 
for interacting many-body systems}
\label{sec2}

We introduce the $k$-body embedded Gaussian ensembles of random matrices for bosons and 
fermions, for the cases of orthogonal ($\beta=1$) and unitary ($\beta=2$) symmetries,
following Ref.~\cite{2003JPA-BW}. We consider a set of $l$ degenerate single-particle 
states $|j\rangle$, with $j=1,2,\dots, l$. The associated creation and annihilation 
operators for fermions are $a^\dagger_j$ and $a_j$, and $b^\dagger_j$ and $b_j$ for bosons, 
with $j=1,\dots, l$. These operators obey the usual (anti)commutation relations that 
characterize the corresponding particles. We define the operators that create a normalized 
state with $k < l$ fermions from the vacuum as 
$\psi^\dagger_{k;\alpha} = \psi^\dagger_{j_1,\dots, j_k} = \prod_{s=1}^k a^\dagger_{j_s}$, 
with the convention that the indexes are ordered increasingly $j_1<j_2<\cdots < j_k$ ($\alpha$
simplifies the notation for these indexes); the corresponding annihilation operators are 
$\psi_{k;\alpha} = (\psi^\dagger_{k;\alpha})^\dagger$. Likewise, the $k$-boson states are 
given by $\chi^\dagger_{k;\alpha} = \chi^\dagger_{j_1,\dots,j_k} = N_\alpha \prod_{s=1}^k 
b^\dagger_{j_s}$, where again $j_1 \leq j_2 \leq \cdots \leq j_k$. Here, $N_\alpha$ 
is a factor that guarantees the normalization to unity of $\chi^\dagger_{k;\alpha}|0\rangle$: if 
the index $j$ repeats $k_j$ times, $N_\alpha$ contains a factor $(k_j!)^{-1/2}$. 

The random $k$-body Hamiltonian for fermions reads
\begin{equation}
\label{eqHam}
H_k^{(\beta)} = \sum_{\alpha, \gamma} v^{(\beta)}_{k; \alpha, \gamma} 
\psi^\dagger_{k;\alpha} \psi_{k;\gamma};
\end{equation}
a similar equation holds for bosons replacing $\psi^\dagger_{k;\alpha}$ by 
$\chi^\dagger_{k;\alpha}$. In Eq.~(\ref{eqHam}) the coefficients 
$v^{(\beta)}_{k; \alpha, \gamma}$ are random distributed independent Gaussian variables with 
zero mean and constant variance 
\begin{equation}
\label{variance}
\overline{v^{(\beta)}_{k; \alpha, \gamma} v^{(\beta)}_{k; \alpha', \gamma'}} = 
v_0^2 (\delta_{\alpha,\gamma'}\delta_{\alpha',\gamma} + 
\delta_{\beta,1}\delta_{\alpha,\alpha'}\delta_{\gamma,\gamma'}).
\end{equation}
Here, $\beta$ is Dyson's parameter that accounts for the presence ($\beta=1$)
or absence ($\beta=2$) of time-reversal 
symmetry~\cite{guhr}, the bar denotes ensemble average, and we set $v_0=1$ without loss
of generality.

The Hamiltonian $H_k^{(\beta)}$ acts on a Hilbert space spanned by distributing $n\geq k$ 
particles on the $l$ single-particle states. Then, a complete set of basis states is given 
by the set $\psi^\dagger_{n;\alpha} |0\rangle$ for fermions (with $l>n$), and 
$\chi^\dagger_{n;\alpha} |0\rangle$
for bosons. The dimension of the Hilbert spaces are, respectively, $N_F=\binom{l}{n}$ and 
$N_B=\binom{l+n-1}{n}$. This defines the $k$-body embedded Gaussian ensembles of random 
matrices~\cite{2003JPA-BW, kotabook}. 

By construction, the case $k=n$ is identical to the canonical ensembles of Random Matrix
Theory~\cite{guhr}, i.e., to the Gaussian Orthogonal Ensemble (GOE) for $\beta=1$ or
the Gaussian Unitary ensemble (GUE) for $\beta=2$. For $k < n$, the matrix elements of 
$H_k^{(\beta)}$ may be identical 
to zero and display correlations. The former property appears whenever there are no 
$k$-body operators that link together the $n$-body states, e.g., for $k< n/2$. Correlations 
arise because matrix elements of $H_k^{(\beta)}$ not related by symmetry may be 
identical. One of the notorious consequences of this is, for the bosonic ensemble 
in the dense limit ($n\gg k,l$, for $k$ and $l$ fixed) that the ensemble is 
non-ergodic~\cite{2002AnnPhys}. In particular, for $l=2$, $k\ll n$ and 
$\beta=1$, the spectrum displays a significant number of quasi-degenerate 
states~\cite{saul}.

As mentioned above, centrosymmetry is an important concept for an optimal 
efficiency~\cite{2014NJP-ZMWB}. A symmetric $N\times N$ matrix $A$ is centrosymmetric if 
$[A,J]=0$, where $J$ is the {\it exchange matrix} 
$J_{i,j} = \delta_{i,N-i+1}$~\cite{cantoni}; 
for complex hermitian matrices, centrohermiticity is defined when 
$J A^T J = A$~\cite{1990SIAMMatAnalApp}.

Imposing centrosymmetry to the $k$-body embedded ensembles is subtle. It can be 
introduced either at the one-particle level, which is the core for the definition
of the $k$- and $n$-particle Hilbert spaces, or at the $k$-body level, 
where the actual (random) parameters of the embedded ensembles are set, or at the $n$-body 
level, where the system evolves. Considering 
a more realistic description which includes a one-body (mean-field) term and a 
two-body (residual) interaction, $H = H_{k=1}+H_{k=2}$, it seems unnatural 
to define a specific transformation for each term separately. Hence, we shall
introduce it in the one-particle space, and compute how is it 
transferred to the $k$-body and $n$-body space, which depends on $\beta$ as well. Then, 
in the one-particle space we define 
$J_1 |j\rangle = |l-j+1\rangle$ for $j=1,\dots, l$, whose matrix representation in the 
one-body basis is precisely the exchange matrix. For 
two fermions, we define 
$J_2 \psi^\dagger_{2;j_1,j_2} = J_1 a^\dagger_{j_1} J_1 a^\dagger_{j_2} = 
- \psi^\dagger_{2;l-j_2+1,l-j_1+1}$. In the last equality we keep the convention that 
the indexes are arranged in increasing order; then, the fermionic anticommutation relations 
impose a global minus sign, which can be safely ignored. This is generalized 
for $k$ particles as
\begin{equation}
\label{CS}
J_k \psi^\dagger_{k;j_1,\dots,j_k} = \prod_{s=1}^k J_1 a^\dagger_{j_s} = 
\psi^\dagger_{k;l-j_k+1,\dots,l-j_1+1}\ ,
\end{equation}
where we have dropped any global minus sign. We note that in general the matrix
$J_k$, as defined 
by Eq.~(\ref{CS}), is not an exchange matrix, i.e., the matrix with ones in the 
counterdiagonal and zeros elsewhere. This follows from the possible existence of more 
than one state that is mapped by $J_k$ onto itself; in this case, we shall say that
$J_k$ is a {\it partial} exchange matrix. As an example, consider the 
case of fermions distributed in $l=4$ single-particle states; for $k=2$ the 
$k$-particle space has dimension $6$. Then, $J_2 \psi^\dagger_{2;2,3} = 
\psi^\dagger_{2;2,3}$ and $J_2 \psi^\dagger_{2;1,4} = \psi^\dagger_{2;1,4}$, 
ignoring the minus signs mentioned above. Then, the entries in the $J_2$ matrix elements 
for these basis states are 1 in the diagonal. In contrast, for the case $l=4$ and $k=1,3$ 
the resulting matrices $J_1$ and $J_3$ are exchange matrices.

For bosons $J_k$ is defined as for fermions using Eq.~~(\ref{CS}) with the
corresponding change of the creation operators. In this case, again, $J_k$ may not be 
an exchange matrix. As an example consider $k=2$ and $l=3$; then, we have 
$J_2 \chi^\dagger_{2;2,2} = N_\alpha (J_1 b^\dagger_{2})^2 = \chi^\dagger_{2;2,2}$,  
and likewise $J_2 \chi^\dagger_{2;1,3} = \chi^\dagger_{2;1,3}$, which shows that
there are more than one basis states mapped onto themselves under $J_2$. Yet, for 
the special case $l=2$ that we study below, $J_k$ is an exchange matrix for all
$k$, and $H_k$ 
(in the $n$-particle space) inherits the centrosymmetry (or centrohermiticity) of 
$v_k$, i.e., $[J_n, H_k]=0$, with $J_n$ the exchange matrix of appropriate 
dimensions.

\section{Statistics of the transport efficiency for bosons distributed in $l=2$ levels}
\label{sec3}

We analyze here the transport efficiency ${\cal P}_{\mu,\nu}$ for bosons distributed 
in $l=2$ single-particle states, considering an arbitrary initial state 
$\chi_{n;\mu}^\dagger$ 
and an arbitrary final one $\chi_{n;\nu}^\dagger$. We address separately the cases where 
the $k$-body random interaction is invariant or not with respect to time-reversal 
($\beta=1$ or $\beta=2$), and also when centrosymmetry (or centrohermiticity) 
is or not additionally imposed. In the simulations described below, we considered 
for concreteness the case with $n=9$ bosons, the corresponding Hilbert space dimension 
is $N_B=10$, the number of realizations of the ensemble is $2000$, and 
$T=15$ in Eq.~(\ref{eq1}).

\begin{figure*}
\centering
\includegraphics[scale=1.25]{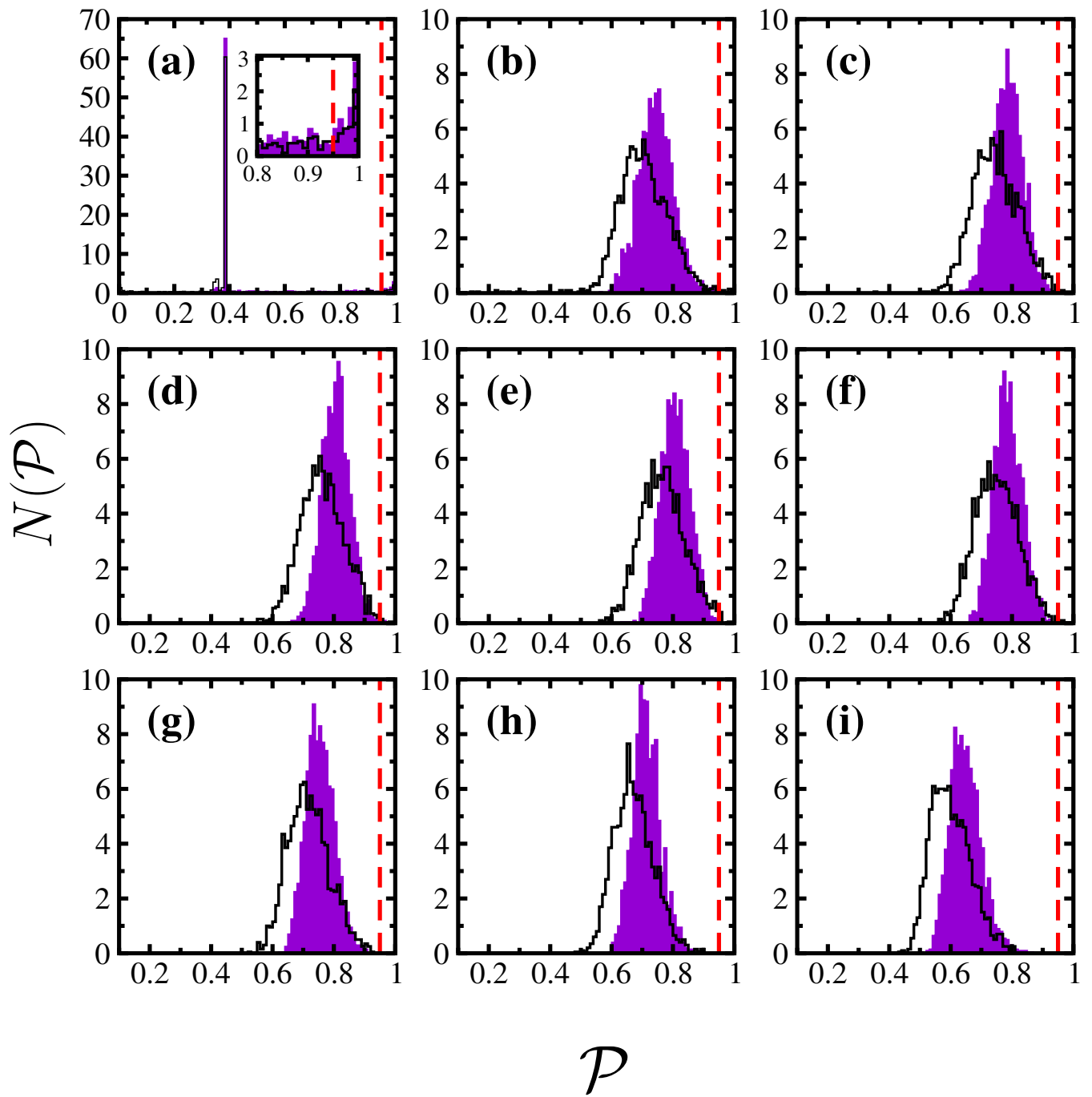}
\caption{%
\label{fig:fig1}
Normalized distributions of the best efficiencies ${\cal P}$ of each realization 
of the bosonic embedded Gaussian ensemble with no centrosymmetry (or 
centrohermiticity) for $l=2$, $n = 9$: (a)~$k=1$, (b)~$k=2$, (c)~$k=3$,
(d)~$k=4$, (e)~$k=5$, (f)~$k=6$, (g)~$k=7$, (h)~$k=8$, and (i)~$k=9$. The empty 
histograms correspond to $\beta = 1$ and the shaded (violet) to $\beta = 2$. The red 
vertical line indicates the 95\% benchmark for the efficiency. Note that the
scales for the $k=1$ case are different from the rest. The inset in (a) is an 
enlargement of a region close to the benchmark value, showing the non-zero probability 
of having PST.}
\end{figure*}

We discuss first the case where centrosymmetry or centrohermiticity, respectively
for $\beta=1$ and $\beta=2$, is absent. The distributions of the efficiencies 
${\cal P}_{\mu,\nu}$ of the ensemble for 
any combination of the input and output states is in general rather broad, 
with negligible contributions to efficiencies above $0.95$. More interesting is 
to focus on the distribution of best efficiencies ${\cal P}$ of each realization 
of the ensemble. The results are presented in Fig.~\ref{fig:fig1}. The empty
histograms correspond to $\beta=1$ and the shaded ones to $\beta=2$. With respect 
to time-reversal invariance, the case $\beta=2$ seems to yield marginally better
efficiencies for a given value of $k$, in
the sense that the mean value of the distribution attains larger values 
and the distribution is somewhat narrower. Yet, we notice that 
the largest values of ${\cal P}$
are slightly dominated by the $\beta=1$ case, except for $k=1$. Regarding the 
dependence on $k$, the case $k=1$ notably distinguishes itself as a special one. 
First, the distribution of ${\cal P}$ is dominated by a peak at $0.4$, is
wider than for other values of $k$, and displays the largest number of realizations
with efficiencies larger than the benchmark value $0.95$ 
(vertical dashed line). Apart from this case, 
other values of $k$ display rather poor efficiencies, with smaller probabilities
of reaching the benchmark value if they do it at all. Actually, the distributions
for intermediate values of $k$ ($4 \leq k \leq 6$) seem to be marginally better. 
In general, the states $\chi_{n;\mu}^\dagger$ 
and $\chi_{n;\nu}^\dagger$ that yield the best efficiencies differ by one boson in the 
occupation of each single-particle state, except for $k=9$ where the best 
efficiency is uniformly distributed for all pairs of states. This statement
is consistent with the fact that 
$k=n$ corresponds to a GOE or GUE. Clearly, these unconstrained embedded Gaussian 
ensembles yield poor transfer efficiencies, except maybe for $k=1$.

\begin{figure*}
\centering
\includegraphics[scale=1.25]{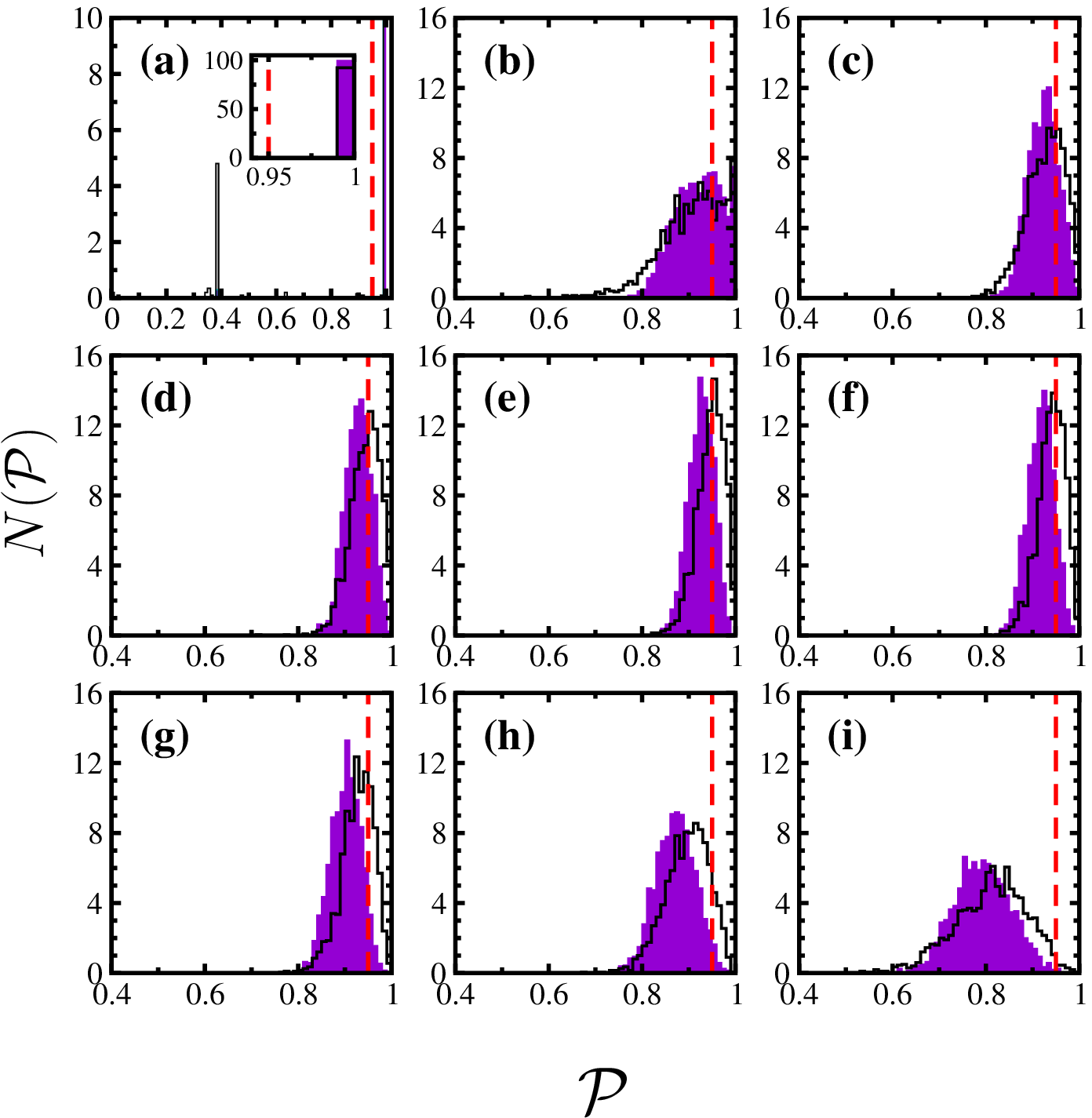}
\caption{%
\label{fig:fig2}
Same as Fig.~\ref{fig:fig1} where we impose centrosymmetry or centrohermiticity 
for the embedded Gaussian ensembles. Note that for $k=1$ and $\beta=2$, almost 
all realizations yield almost PST.}
\end{figure*}

We turn now to the best efficiencies of each realization of the ensemble when 
centrosymmetry or centrohermiticity is imposed.
As mentioned above, for the specific bosonic ensemble with $l=2$, imposing
centrosymmetry or centrohermiticity at the one-particle level carries over to 
the $k$- and $n$-particle
spaces; hence, $[H_k,J_n]=0$, with $J_n$ the $N_B\times N_B$ exchange matrix.
In Fig.~\ref{fig:fig2} we present the best efficiencies ${\cal P}$ of each 
realization of the ensemble for this case. In comparison to the results presented
in Fig.~\ref{fig:fig1}, centrosymmetry (centrohermiticity) enhances
dramatically the best efficiencies. In particular, the case 
$k=1$ is remarkable since it displays perfect state transfer 
for {\it all} realizations of the ensemble for $\beta=2$, while for 
$\beta=1$ there is a small remnant of a peak around $0.4$. Quantitatively, for 
$\beta=1$ about $92\%$ of the realizations exhibit efficiencies that are larger 
than the benchmark value, while for $\beta=2$ this number is well above $99\%$; 
the mean of the distributions is $0.9556$ and $0.9978$, respectively. Increasing 
$k$ shifts the mean value of the distribution towards smaller values and widens
the distributions, being the effect larger for $\beta=2$ than for $\beta=1$. 
Whereas for $k=2$ some realizations still display almost PST (last bin of the 
histograms), this decreases for increasing values of $k$. For $\beta=1$,
all values of $k$ exhibit realizations whose best efficiencies are above the 
benchmark value, with better efficiencies appearing for smaller values of $k$. 

The states $\chi_{n;\mu}^\dagger$ and $\chi_{n;\nu}^\dagger$ that display 
the best efficiencies are those linked by centrosymmetry, i.e., 
$J_n \chi_{n;\mu}^\dagger = \chi_{n;\nu}^\dagger$. While for $k=n$ all 
centrosymmetry-related pairs participate uniformly, for other
values of $k$ there is a significant dominance of the pair of states 
where all bosons are initially in either of the single-particle states. From
the perspective of the network, these states are located at the edges.

\begin{figure*}
\centering
\includegraphics[scale=1.0]{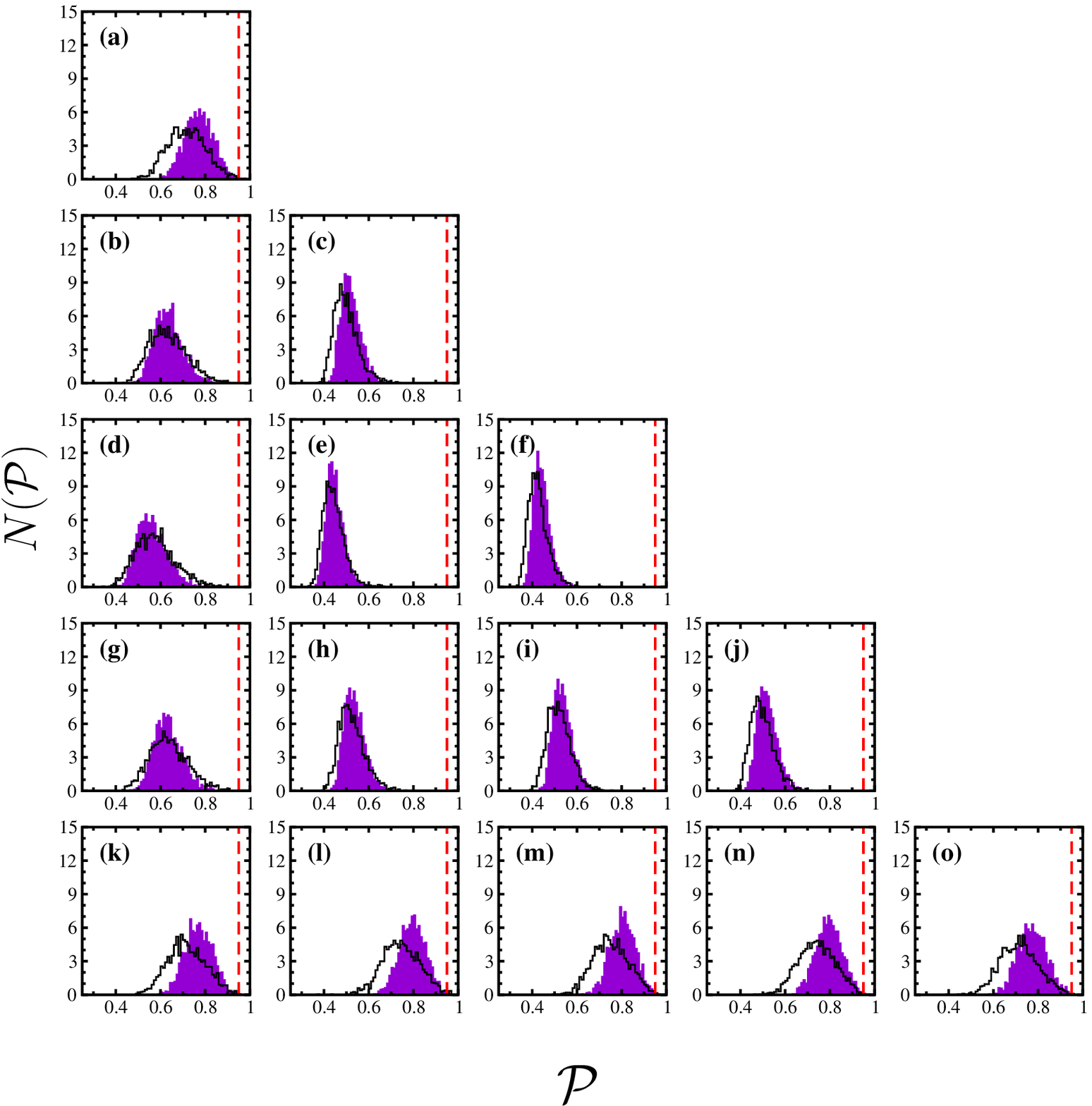}
\caption{
\label{fig:fig3}
Normalized distributions of the best efficiencies ${\cal P}$ of each realization 
of the the fermionic embedded Gaussian ensemble without centrosymmetry or 
centrohermiticity for $l=6$: (a)~$k=n=1$, 
(b)~$k=1, n=2$, (c)~$k=2, n=2$, 
(d)~$k=1, n=3$, (e)~$k=2, n=3$, (f)~$k=n=3$,
(g)~$k=1, n=4$, (h)~$k=2, n=4$, (i)~$k=3, n=4$, (j)~$k=n=4$
(k)~$k=1, n=5$, (l)~$k=2, n=5$, (m)~$k=3, n=5$, (n)~$k=4, n=5$, (o)~$k=n=5$.
(Rows have the same particle number $n$, and columns the same $k$ value). Note that the 
two last rows correspond to filling factors above half-filling. Empty histograms
correspond to $\beta=1$ and shaded (violet) to $\beta=2$.}
\end{figure*}

As mentioned above and manifested in Fig.~\ref{fig:fig2}(a), the case 
$k=1$ is quite special. Indeed, for $k=1$ there are only three ($\beta=1$) 
and four ($\beta=2$) independent random matrix elements; centrosymmetry 
(centrohermiticity) imposes that they become only two ($\beta=1$) and three 
($\beta=2$). The consequence of this is that the random many-body 
centrosymmetric/centrohermitian Hamiltonian has a constant main diagonal, which 
is proportional to the total number of bosons, and a second-diagonal whose matrix 
elements are proportional to $\sqrt{(n_i+1)(N_B-n_i-1)}$, 
where $n_i=0,\dots, n$ is the boson occupation number of one of the single-particle 
levels for the many-body state in question. 
Except for the constant diagonal and the random weights involved, for $\beta=1$ 
this matrix model has been 
shown to exhibit PST~\cite{2006PRL-F}. It can be shown that
the eigenvectors of both models are identical, and the eigenvalues are related 
by a linear function. This ensures the occurrence of PST
as long as the off-diagonal matrix element of the random one-body interaction matrix 
is different from zero. Yet, the time scale where the PST occurs may be quite
long, since it is inversely proportional to the absolute value of the off-diagonal
matrix element of the random one-body matrix; this explains the occurrence of
other values for the efficiency, as shown in Fig.~\ref{fig:fig2}(a). Our results 
extend the validity of 
these statement to the broken time-reversal symmetric case and, more important, 
illustrate that they are robust under certain random centrosymmetric/centrohermitian 
perturbations that preserve the graph structure. We emphasize the important role 
of centrosymmetry or centrohermiticity here: it guarantees that the diagonal 
matrix elements have a constant value, which results in PST among the edge
states of the network.

\section{Statistics of the transport efficiency for fermions}
\label{sec4}

\begin{figure*}
\centering
\includegraphics[scale=1.0]{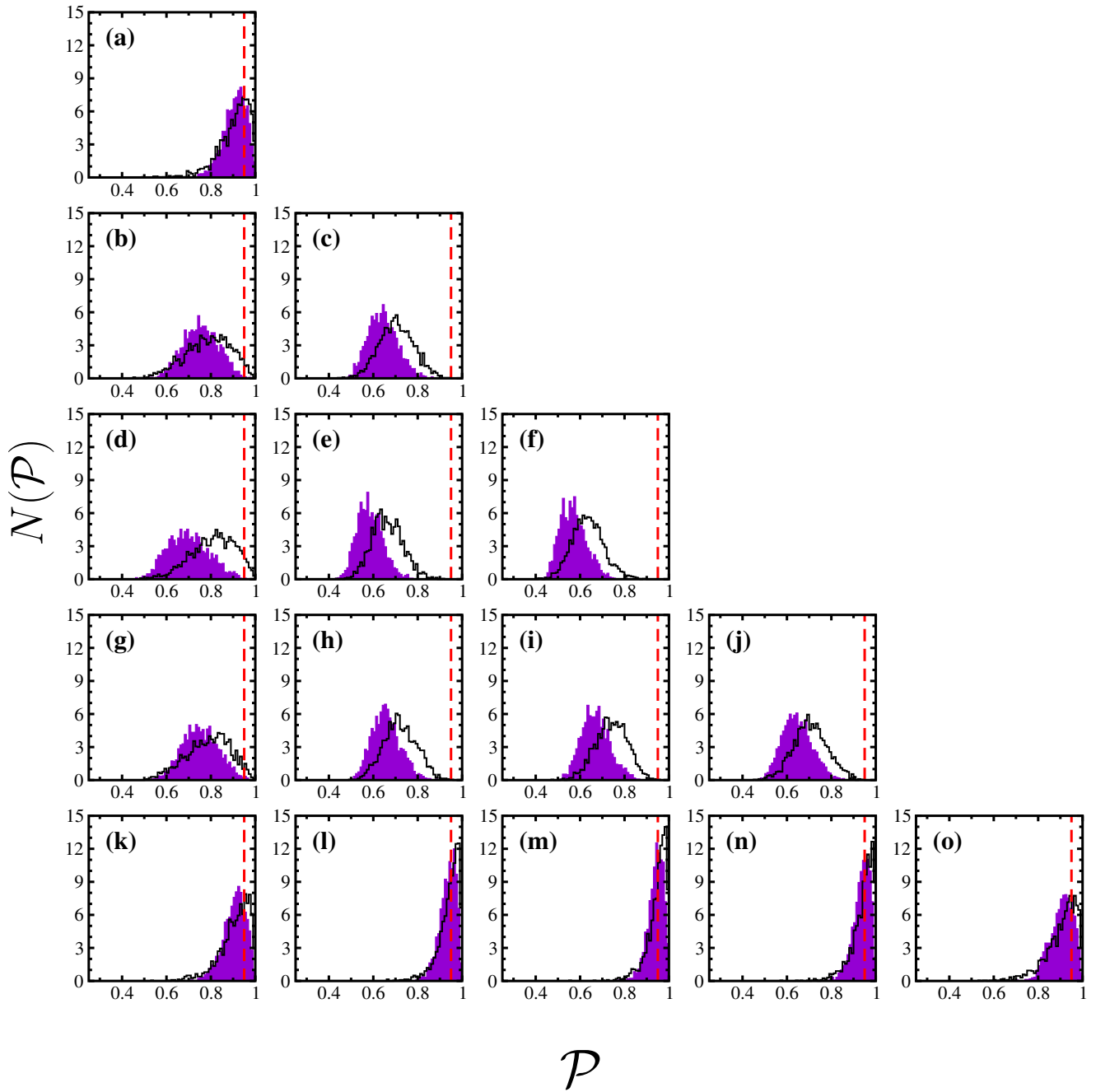}
\caption{
\label{fig:fig4}
Same as Fig.~\ref{fig:fig3} where we impose centrosymmetry or centrohermiticity
in the one-particle space. Its extension to the $k$- and $n$-particle spaces
is such that full centrosymmetry/centrohermiticity is achieved for odd values 
of $k$ or $n$. Note that combining full centrosymmetry/centrohermiticity 
and being above half-filling, for $k\sim l/2$ yields the best efficiencies
per realization.}
\end{figure*}

We consider now the case of the fermionic embedded Gaussian ensemble. For 
concreteness we analyze the case $l=6$, varying the number of fermions $n$ from 
$1$ to $5$, and rank of interaction $k$ also from $1$ to $5$. In general, the
distribution of efficiencies of the ensemble is rather broad with negligible
contribution to efficiencies above $95$\% and thus, we will focus on distribution of 
best efficiencies ${\cal P}$ for each member of the ensemble. 

We begin describing first the case where no centrosymmetry or centrohermiticity 
is considered. The distribution of the best efficiencies $\mathcal{P}$ of each 
realization are 
illustrated in Fig~\ref{fig:fig3}. In this figure, all frames in a row have 
the same number of 
fermions $n$, and all frames in a column the same value of $k$. As before, the empty histograms 
correspond to $\beta=1$ and the filled ones to $\beta=2$. In general, the 
lack of centrosymmetry or centrohermiticity yields poor efficiencies, with a 
marginal improvement for $\beta=2$ compared to the $\beta=1$ case. Interestingly, 
below and at half-filling (first three rows) only the case $k=n=1$ marginally reaches the 
benchmark value, while above half-filling (last two rows of the figure) the 
distributions of $\mathcal{P}$ yield better results. In particular, the case
$n=5=l-1$ is the only one where some realizations achieve the benchmark value. 
Yet, the probability of such events is very small.

The distributions of the best efficiencies of each realization including 
centrosymmetry/centrohermiticity are displayed in Fig.~\ref{fig:fig4}. As in 
the bosonic case, $\beta=1$ seems to yield better efficiencies than the 
$\beta=2$ case. For $k=1$, the distributions of ${\cal P}$ for all $n$ surpass 
the benchmark value, and have a non-zero probability
of displaying PST, specially for odd fermion number. Yet, the $k=1$ case 
does not correspond to the overall optimal situation: the probability that ${\cal P}$ 
is above the benchmark value is clearly larger for $n=5=l-1$ and $2\leq k \leq 4$. 
Indeed, while we find that nearly $30\%$ of the best efficiencies are larger 
than the benchmark for $k=n=1$ and $\beta=1$ (similar values are obtained for
$k=1$ and $k=5$ for $n=5$), for $k=3$ and $n=5$ this percentage is close to $59\%$.
For $\beta=2$ similar results are found with smaller percentages.

While centrosymmetry clearly enhances the efficiencies, its role is more involved. As 
we mentioned above, centrosymmetry/centrohermiticity is introduced at the 
one-particle level and then extended onto the $k$- and $n$-particle spaces
using Eq.~\ref{CS}. The corresponding $J_k$ and $J_n$ matrices are exchange matrices 
or partial exchange matrices depending on the parameters $k$ and $n$, and $l$. For
the present case, $l=6$, it can be shown that for odd particle-number space ($k$ 
or $n$) the corresponding matrix ($J_k$ or $J_n$) is a {\it bona fide} exchange 
matrix, whereas for even-values there are more than one states that are mapped 
onto itself, i.e., the corresponding $J_k$ or $J_n$ matrix is a partial exchange 
matrix. Therefore, independent of $k$, including the case where $J_k$ is a partial 
exchange matrix, the $n=3$ and $n=5$ many-body Hamiltonians are fully 
centrosymmetric/centrohermitian. As shown in Fig.~\ref{fig:fig4}, full 
centrosymmetry or centrohermiticity, in comparison with the partial cases, yield 
the best efficiency distributions. When the many-body Hamiltonian is partial 
centrosymmetric/centrohermitian, the best distributions correspond to the $k=1$ 
case. In addition to centrosymmetry/centro\-hermiticity, the filling factor 
appears to play an important role: above half-filling the distributions 
of ${\cal P}$ display the largest probability to yield efficiencies above the 
benchmark value; in terms of $k$, the optimal case corresponds to $k\simeq l/2$.

Regarding the pair of states that display the best efficiencies when 
centrosymmetry/centrohermiticity is imposed, as in the bosonic case, these occur 
among centrosymmetry related pairs, $\psi_{n;\nu}^\dagger = J_n \psi_{n;\mu}^\dagger$, 
excluding those which are mapped onto themselves by $J_n$ whenever $J_n$ is partial 
centrosymmetric. These pairs of states appear 
uniformly distributed, that is, there is no special pair of states that
display better transport properties. This may be related to the fact that 
the fermionic graph is regular~\cite{2001AnnPhys-BRW}. 

We have confirmed these results for $l=7$, $n=6$, $k=1$, $\dots$, $6$ and $l=8$, 
$n=7$, $k=1$, $\dots$, $7$.

\section{Conclusions}
\label{sec5}

The primary aim of the present paper is to study transport of excitations in
disordered networks with random $k$-body interactions. This
is important and certainly of interest because of possible applications in
quantum communication protocols~\cite{qnet_book} and artificial solar 
cells~\cite{Darius-book}.
Towards this end, we have studied the distribution of quantum efficiencies in
disordered networks with many-body interactions, whose structure is modeled
by the embedded Gaussian ensemble, considering bosons and fermions, with and
without time-reversal symmetry. In particular, we studied the role played by
centrosymmetry/centro\-hermiticity, which is defined at the one-particle space,
and then extended to the $k$- and $n$-particle spaces. We have
shown that (full) centrosymmetry enhances the efficiencies dramatically, being
a requirement to have non-zero probability for PST; the
lack of centrosymmetry/centrohermiticity yields rather poor efficiencies.

For bosons distributed in two single-particle levels, centrosymmetry is inherited
in the $k$- and $n$-body spaces. In this case, PST is obtained for $k=1$
for almost all realizations; the fact that our computation of the efficiencies
involves an upper bound for the time, $T$, constrains the relevant time
scale for the achievement of the PST. In terms of $k$, the probability of having 
PST decays with increasing $k$. However, we stress that for $k>1$, the best efficient 
scenario is when $k\sim n/2$, $n$ is the total number of bosons. With respect to 
the value of $\beta$, the results are marginally better when time-reversal 
symmetry is preserved. The pairs of states showing the best efficiencies are those 
at the edges of the network, i.e., where all bosons are in one of the two 
single-particle levels. Then, in this case, state transfer corresponds to the
physical transport of all bosons to the other single-particle state.

For fermions, we found that full centrosymmetry/cen\-tro\-hermiticity of the 
$n$-particle Hamiltonian enhances considerably the best efficiencies, especially
when the filling-factor is larger than $1/2$. We note that centrosymmetry/centrohermiticity
is indroduced at the one-particle level, and then extended to the $n$-body space.
For fermions, the rank of the 
interaction which displays the highest probability that the best efficiency 
${\cal P}$ is above the benchmark value corresponds to $k\simeq l/2$. A clear 
explanation of this is left open. The pairs of 
states that yield the best efficiencies appear uniformly among those states
linked by centrosymmetry.

Previous results have shown that random perturbations on networks with PST destroy 
or affect significantly this property; for details see~\cite{chiara, casaccino} and 
also~\cite{qnet_book} and references therein. Our results show that, despite of
the random character of the $k$-body interactions that we have considered, certain 
$n$-body networks display good efficiencies and may attain near perfect state transfer with non-zero probability. 

Our results could be exploited as new design principles of networks 
with good efficiency, which is preserved under certain many-body random perturbations.
For instance, considering the implementation of efficient quantum wires, it may be 
interesting to consider the case of filling factors that are smaller but close 
to one, where many-body interactions yield robustly very good efficiencies. Finally, 
the results in our paper open the possibility to understand the good efficiency properties
experimentally observed in exciton transport in biological 
systems, such 
as the Fenna-Matthews-Olson complex~\cite{lee_science,fennamatthews_nature,Darius-book}.

\begin{acknowledgements}
  It is a pleasure to thank Thomas Gorin, Mattia Walschaers and Andreas Buchleitner for 
enlightening discussions and correspondence. We thank the financial support provided 
by the DGAPA-UNAM project IG-101113. MV is DGAPA-UNAM postdoctoral fellow. LB 
acknowledges support through a C\'atedra Marcos Moshinsky (2012).

\end{acknowledgements}


\providecommand{\WileyBibTextsc}{}
\let\textsc\WileyBibTextsc
\providecommand{\othercit}{}
\providecommand{\jr}[1]{#1}
\providecommand{\etal}{~et~al.}

\end{document}